# The nuclear fusion reaction rate based on relativistic equilibrium velocity distribution

Jian-Miin Liu*
Department of Physics, Nanjing University
Nanjing, The People's Republic of China
*On leave. E-mail address: liu@phys.uri.edu

ABSTRACT
   The Coulomb barrier is in general much higher than thermal energy. Nuclear fusion reactions occur only among few protons and nuclei with higher relative energies than Coulomb barrier. It is the equilibrium velocity distribution of these high-energy protons and nuclei that participates in determining the rate of nuclear fusion reactions. In the circumstance it is inappropriate to use the Maxwellian velocity distribution for calculating the nuclear fusion reaction rate. We use the relativistic equilibrium velocity distribution for this purpose. The rate based on the relativistic equilibrium velocity distribution has a reduction factor with respect to that based on the Maxwellian distribution, which factor depends on the temperature, reduced mass and atomic numbers of the studied nuclear fusion reactions. This signifies much to the solar neutrino problem.

**Introduction**.   To create a nuclear fusion reaction, a proton or nucleus must penetrate the repulsive Coulomb barrier and be close to another proton or nucleus so that the strong interaction force between them acts. The Coulomb barrier is in general much higher than thermal energy. Nuclear fusion reactions can take place only among few protons and nuclei with higher relative energies than Coulomb barrier. If, under the conditions for nuclear fusion reactions, interacting protons and nuclei reach their equilibrium distribution in the period of time that is infinitesimal compared to the mean lifetime of nuclear fusion reactions, it is the equilibrium velocity distribution of these high-energy protons and nuclei that participates in determining the rate of nuclear fusion reactions. In this circumstance, it is inappropriate to use the Maxwellian velocity distribution to calculate the nuclear fusion reaction rate.
   In our recent work, we made the relativistic corrections to the Maxwellian velocity distribution [1,2] and explained the observed non-Maxwellian high-energy tails in velocity distributions of astrophysical plasma particles [3]. We name the corrected distribution the relativistic equilibrium velocity distribution keeping in mind that the question whether such a name is suited to it is a matter to be decided in the light of experience. In this letter, we use the relativistic equilibrium velocity distribution to calculate the nuclear fusion reaction rate.

**Velocity space and relativistic equilibrium velocity distribution**.   Some results in Refs.[1,2] are quoted here for use below. The velocity space in the special theory of relativity and the modified special relativity theory [4] can be represented by

$$dY^2 = \delta_{rs} dy'^r dy'^s, \quad r,s=1,2,3, \quad (1)$$

in the so-called primed velocity-coordinates $\{y'^r\}$, r=1,2,3, or by

$$dY^2 = H_{rs}(y) dy^r dy^s, \quad r,s=1,2,3, \quad (2a)$$
$$H_{rs}(y) = c^2 \delta^{rs}/(c^2-y^2) + c^2 y^r y^s/(c^2-y^2)^2, \quad \text{real } y^r \text{ and } y<c, \quad (2b)$$

in the usual velocity-coordinates $\{y^r\}$, r=1,2,3, where $y'^r$ is called the primed velocity, $y^r$ is the well-defined Newtonian velocity, $y=(y^r y^r)^{1/2}$ is the radial component of $y^r$ that we simply call the radial velocity, and c is the speed of light. The primed and the usual velocity-coordinates are connected by

$$dy'^r = A^r_s(y) dy^s, \quad r,s=1,2,3, \quad (3a)$$
$$A^r_s(y) = \gamma \delta^{rs} + \gamma(\gamma-1) y^r y^s/y^2, \quad (3b)$$

because $\delta_{rs} A^r_p(y) A^s_q(y) = H_{pq}(y)$, where

$$\gamma = 1/(1-y^2/c^2)^{1/2}. \quad (4)$$

In the velocity space, the velocity-length between two primed velocities $y'^r_1$ and $y'^r_2$ or two Newtonian velocities $y^r_1$ and $y^r_2$ is

$$Y(y'^r_1, y'^r_2) = [(y'^r_2 - y'^r_1)(y'^r_2 - y'^r_1)]^{1/2}, \quad (5)$$

or

$$Y(y_1^r, y_2^r) = \frac{c}{2} \ell n \frac{b+a}{b-a}, \tag{6a}$$

$$b = c^2 - y_1^r y_2^r, \quad r = 1,2,3, \tag{6b}$$

$$a = \{(c^2 - y_1^i y_1^i)(y_2^j - y_1^j)(y_2^j - y_1^j) + [y_1^k(y_2^k - y_1^k)]^2\}^{1/2}, \quad i,j,k = 1,2,3. \tag{6c}$$

Eqs.(5) and (6a-6c) imply

$$y'^r = [\frac{c}{2y} \ell n \frac{c+y}{c-y}] y^r, \quad r = 1,2,3, \tag{7a}$$

$$y' = \frac{c}{2} \ell n \frac{c+y}{c-y}, \tag{7b}$$

when $(y'^1, y'^2, y'^3)$ and $(y^1, y^2, y^3)$ represent the same point, where $y' = (y'^r y'^r)^{1/2}$. Differentiating Eq.(7b), we get

$$dy' = \frac{dy}{(1 - y^2/c^2)}. \tag{8}$$

In the velocity space, the Galilean addition law of primed velocities links up to the Einstein addition law of their Newtonian velocities. The Euclidean structure of the velocity space in the primed velocity-coordinates convinces us of the Maxwellian distribution of primed velocities,

$$P(y'^1, y'^2, y'^3) dy'^1 dy'^2 dy'^3 = N(\frac{m}{2\pi K_B T})^{3/2} \exp[-\frac{m}{2K_B T}(y')^2] dy'^1 dy'^2 dy'^3, \tag{9a}$$

$$P(y') dy' = 4\pi N(\frac{m}{2\pi K_B T})^{3/2} (y')^2 \exp[-\frac{m}{2K_B T}(y')^2] dy'. \tag{9b}$$

where N is the number of particles, m their rest mass, T the temperature, and $K_B$ the Boltzmann constant. Putting Eqs.(3a-3b), (7b) and (8) into Eqs.(9a-9b), we find the relativistic equilibrium distribution of Newtonian velocities,

$$P(y^1, y^2, y^3) dy^1 dy^2 dy^3 = N \frac{(m/2\pi K_B T)^{3/2}}{(1 - y^2/c^2)^2} \exp[-\frac{mc^2}{8K_B T}(\ell n \frac{c+y}{c-y})^2] dy^1 dy^2 dy^3, \tag{10a}$$

$$P(y) dy = \pi c^2 N \frac{(m/2\pi K_B T)^{3/2}}{(1 - y^2/c^2)} (\ell n \frac{c+y}{c-y})^2 \exp[-\frac{mc^2}{8K_B T}(\ell n \frac{c+y}{c-y})^2] dy. \tag{10b}$$

This distribution is close-fitting to the Maxwellian distribution for low-energy particles (y<<c) but substantially differs from the Maxwellian distribution for high-energy particles. As y goes to c, it falls off to zero slower than any exponential decay and faster than any power-law decay [3].

**Transformation of relative velocities.** Let particle 1 and particle 2 move with Newtonian velocities $y_1^r$ and $y_2^r$, r=1,2,3, respectively. We denote the relative velocity of particle 2 to particle 1 with $v^r$, r=1,2,3. If three corresponding primed velocities are respectively $y_1'^r$, $y_2'^r$, $v'^r$, r=1,2,3, the Galilean addition law among them reads

$$v'^r = y'^r_2 - y'^r_1, \quad r = 1,2,3, \tag{11}$$

and the Einstein addition law among the three Newtonian velocities is

$$v^r = \sqrt{1 - y_1^2/c^2} \{(y_2^r - y_1^r) + (\frac{1}{\sqrt{1 - y_1^2/c^2}} - 1) y_1^r \frac{y_1^s(y_2^s - y_1^s)}{y_1^2}\} / [1 - \frac{y_2^k y_1^k}{c^2}], \quad r,s,k = 1,2,3. \tag{12}$$

Due to Eqs.(5) and (6a-6c), we have

$$[(y'_2{}^r - y'_1{}^r)(y'_2{}^r - y'_1{}^r)]^{1/2} = \frac{c}{2} \ln \frac{b+a}{b-a}$$

with

$b = c^2 - y_2{}^r y_1{}^r$, $r = 1,2,3$,
$a = \{(c^2 - y_1{}^i y_1{}^i)(y_2{}^j - y_1{}^j)(y_2{}^j - y_1{}^j) + [y_1{}^k(y_2{}^k - y_1{}^k)]^2\}^{1/2}$, $i,j,k = 1,2,3$,

or equivalently

$$c^2 \tanh^2\{[(y'_2{}^r - y'_1{}^r)(y'_2{}^r - y'_1{}^r)]^{1/2}/c\} = c^2\{(c^2 - y_1{}^i y_1{}^i)(y_2{}^j - y_1{}^j)(y_2{}^j - y_1{}^j) + [y_1{}^k(y_2{}^k - y_1{}^k)]^2\}/(c^2 - y_1{}^r y_1{}^r)^2. \quad (13)$$

We separately use Eqs.(11) and (12) on the left-hand and the right-hand sides of Eq.(13) and obtain

$$c^2 \tanh^2\{v'/c\} = v^2, \quad (14)$$

where $v' = (v'^r v'^r)^{1/2}$ and $v = (v^r v^r)^{1/2}$. Eq.(14) specifies the transformation between relative Newtonian velocity $v^r$ and its relative primed velocity $v'^r$,

$$v' = \frac{c}{2} \ln \frac{c+v}{c-v}, \quad (15a)$$

$$v'^r = \left(\frac{c}{2v} \ln \frac{c+v}{c-v}\right) v^r, \quad r = 1,2,3. \quad (15b)$$

Differentiating Eq.(15a) immediately yields

$$dv' = \frac{dv}{(1 - v^2/c^2)}. \quad (16)$$

**Equilibrium distribution of relative velocities.** Suppose we have two types of particles, type 1 and type 2, and the type i particles have mass $m_i$, $i = 1,2$. It has been proved [5] that when velocities of the type 1 particles, as well as those of the type 2 particles, obey the Maxwellian distribution, the relative velocities of the type 2 particles to the type 1 particles obey the like, in which provided we take reduced mass

$\mu = m_1 m_2/(m_1 + m_2)$.

In Eqs.(9a-9b) we see that the primed velocities of the type 1 particles and of the type 2 particles obey the Maxwellian distribution. The distribution of the relative primed velocities of the type 2 particles to the type 1 particles must be in accordance with

$$P^*(v'^1, v'^2, v'^3) dv'^1 dv'^2 dv'^3 = N\left(\frac{\mu}{2\pi K_B T}\right)^{3/2} \exp\left[-\frac{\mu}{2K_B T}(v')^2\right] dv'^1 dv'^2 dv'^3, \quad (17a)$$

$$P^*(v') dv' = 4\pi N\left(\frac{\mu}{2\pi K_B T}\right)^{3/2} (v'^2) \exp\left[-\frac{\mu}{2K_B T}(v'^2)\right] dv'. \quad (17b)$$

Inserting Eqs.(15a) and (16) in Eq.(17b), we obtain the relativistic equilibrium distribution for radial relative (Newtonian) velocities of the type 2 particles to the type 1 particles,

$$P^*(v) dv = \pi c^2 N \frac{(\mu/2\pi K_B T)^{3/2}}{(1 - v^2/c^2)} \left(\ln \frac{c+v}{c-v}\right)^2 \exp\left[-\frac{\mu c^2}{8 K_B T} \left(\ln \frac{c+v}{c-v}\right)^2\right] dv. \quad (18)$$

**Nuclear cross section.** Now we consider a kind of nuclear fusion reactions occurring between protons or nuclei of type 1 and protons or nuclei of type 2, where protons or nuclei of type i have density $N_i$, mass $m_i$ and atomic number $z_i$, $i = 1,2$. The rate of these nuclear fusion reactions is

$$R = \frac{1}{(1+\delta_{12})} N_1 N_2 \langle v \sigma(v) \rangle = \frac{1}{(1+\delta_{12})} N_1 N_2 \int_0^c v \sigma(v) f(v) dv, \quad (19)$$

where v denotes the radial relative velocities of the type 2 protons or nuclei to the type 1 ones, $\sigma(v)$ is a cross section of nuclear fusion reactions of the kind, $\langle \ \rangle$ means the thermodynamic-equilibrium average over v, $0 \leq v < c$. The normalized equilibrium distribution f(v), $\int_0^c f(v) dv = 1$, can be picked out in Eq.(18).

As for the cross section σ(v), its recognized form of $\frac{S}{\mu v^2/2}\exp[-\frac{2\pi z_1 z_2 e^2}{\hbar v}]$ is no longer suitable here.

The cross section σ(v) is a product of three factors: the penetration probability factor, the slowly varying factor S and the factor of the squared de Broglie wavelength. The penetration probability factor describes a distribution of the probability for an incoming proton or nucleus with atomic number $z_1$ to penetrate through the repulsive Coulomb barrier of a target proton or nucleus with atomic number $z_2$ due to quantum tunnel effect [6,7], as well as f(v) is a distribution of the probability for the incoming proton or nucleus to have radial velocity v relative to the target proton or nucleus. The penetration probability factor is also a function of radial relative velocity. We ought to treat it in the same way that we did for f(v). In other words, we take $\exp[-\frac{2\pi z_1 z_2 e^2}{\hbar v'}]$ for the penetration probability factor in the primed velocity-coordinates and $\exp[-2\pi z_1 z_2 e^2/\hbar\,(\frac{c}{2}\ell n\frac{c+v}{c-v})]$ in the usual velocity-coordinates, where Eq.(15a) is used. The other two factors are together with the penetration probability factor to form the cross section as a whole, so we deal with these two factors consistently keeping their original forms in the primed velocity-coordinates and looking for their forms in the usual velocity-coordinates with the aid of Eq.(15a). Totally, for the cross section, we have

$$\frac{S}{\mu v'^2/2}\exp[-\frac{2\pi z_1 z_2 e^2}{\hbar v'}] \qquad (20)$$

in the primed velocity-coordinates and

$$\{S/\frac{\mu}{2}(\frac{c}{2}\ell n\frac{c+v}{c-v})^2\}\exp[-2\pi z_1 z_2 e^2/\hbar\,(\frac{c}{2}\ell n\frac{c+v}{c-v})] \qquad (21)$$

in the usual velocity-coordinates. Readers may refer to Ref.[8] for more explanations of this treatment.

**Nuclear fusion reaction rate**.   To calculate the nuclear fusion reaction rate, we take P*(v)/N in Eq.(18) for f(v) and put it and Eq.(21) into Eq.(19). Introducing a new variable, $x=\frac{c}{2}\ell n\frac{c+v}{c-v}$, we find

$$R=\frac{8\pi}{\mu}(\frac{\mu}{2\pi K_B T})^{3/2}\frac{1}{(1+\delta_{12})}N_1 N_2\int_0^\infty Sc[\tanh(\frac{x}{c})]\exp[-\frac{\mu x^2}{2K_B T}-\frac{2\pi z_1 z_2 e^2}{\hbar x}]dx. \qquad (22)$$

Calling

$$\tanh(\frac{x}{c})=\sum_{n=1}^\infty(-1)^{n+1}\frac{2^{2n}(2^{2n}-1)B_n}{(2n)!}(\frac{x}{c})^{2n-1},$$

in a region, where $B_n$, n=1,2,3,------, are the Bernoulli numbers, $B_1=1/6$, $B_2=1/30$, $B_3=1/42$, ------, and using the method of the steepest descents, we further find

$$R=\sum_{n=1}^\infty R_n, \qquad (23)$$

$$R_n=\frac{1}{(1+\delta_{12})}N_1 N_2\frac{4c}{\sqrt{3}K_B T}S_{eff}\exp[-\lambda](-1)^{n+1}\frac{2^{2n}(2^{2n}-1)B_n}{(2n)!}(2\pi z_1 z_2\frac{K_B T}{\mu c^2}\frac{e^2}{\hbar c})^{\frac{2n-1}{3}},$$

$$(24)$$

where $S_{eff}$ is the average of slowly varying function S and

$$\lambda=\frac{3}{K_B T}(\frac{\sqrt{\mu K_B T}\pi z_1 z_2 e^2}{\sqrt{2}\hbar})^{2/3}$$

is n-independent. The most effective energy is

$$E_0 = \left(\frac{\sqrt{\mu}K_B T \pi z_1 z_2 e^2}{\sqrt{2}\hbar}\right)^{2/3},$$

which is also n-independent. In a compact form, we can rewrite R as

$$R = \frac{1}{(1+\delta_{12})} N_1 N_2 \frac{4c}{\sqrt{3}K_B T} S_{\text{eff}} \exp[-\lambda] \tanh\left\{\left(2\pi z_1 z_2 \frac{K_B T}{\mu c^2} \frac{e^2}{\hbar c}\right)^{1/3}\right\} \qquad (25)$$

or

$$R = \frac{\tanh Q}{Q} R_M, \qquad (26a)$$

$$Q = \left(2\pi z_1 z_2 \frac{K_B T}{\mu c^2} \frac{e^2}{\hbar c}\right)^{1/3}, \qquad (26b)$$

$$R_M = \frac{1}{(1+\delta_{12})} N_1 N_2 \frac{4c}{\sqrt{3}K_B T} S_{\text{eff}} \exp[-\lambda] \left(2\pi z_1 z_2 \frac{K_B T}{\mu c^2} \frac{e^2}{\hbar c}\right)^{1/3}, \qquad (27)$$

where $R_M$ is the nuclear fusion reaction rate based on the Maxwellian velocity distribution.

**Concluding remarks.** The nuclear fusion reaction rate based on the relativistic equilibrium velocity distribution has a reduction factor with respect to that based on the Maxwellian velocity distribution: $\tanh Q/Q$. Since $0<Q<\infty$, the reduction factor satisfies $0<\tanh Q/Q<1$. That gives rise to

$$0 < R < R_M. \qquad (28)$$

The reduction factor depends on the temperature, reduced mass and atomic numbers of the studied nuclear fusion reactions.

For a statistical calculation relevant to equilibrium velocity distribution, in two cases we have to substitute the relativistic equilibrium velocity distribution for the Maxwellian distribution. One case is where most particles crowd in the high-energy region. The other case is in which most particles crowd in the low-energy region but these particles in the low-energy region are not involved in the statistical calculation. The calculation for the nuclear fusion reaction rate belongs here. When most particles crowd in the low-energy region and, concurrently, these particles in the low-energy region are involved in the statistical calculation, substitution for the Maxwellian distribution is not so important.

In solar interior, the height of Coulomb barrier is far above solar thermal energy: their ratio is typically greater than a thousand [9,10], and the interacting protons and nuclei reach their equilibrium distribution in a very short time which is infinitesimal compared to the mean lifetime of a nuclear fusion reaction [10]. The obtained results in the context are applicable to all kinds of solar nuclear fusion reactions. As seen in Eq.(28), the relativistic equilibrium velocity distribution lowers the rates of solar nuclear fusion reactions. It will hence lower the solar neutrino fluxes. It will also change the solar neutrino energy spectra. On the other hand, since most solar ions crowd in the low-energy region, at temperatures and densities in the Sun, and since these ions are involved in the calculation of solar sound speeds, substituting the relativistic equilibrium velocity distribution for the Maxwellian velocity distribution is not important to the calculation of solar sound speeds. The relativistic equilibrium velocity distribution, once adopted in standard solar models, will lower solar neutrino fluxes and change solar neutrino energy spectra but maintain solar sound speeds.

This work does not include any screening correction to the nuclear fusion reaction rate [11-13].

ACKNOWLEDGMENT
The author greatly appreciates the teachings of Prof. Wo-Te Shen. The author thanks Prof. Gerhard Muller for useful suggestions.

REFERENCES
[1]     Jian-Miin Liu, Chaos Solitons&Fractals, 12, 2149 (2001)

[2]     Jian-Miin Liu, cond-mat/0108356
[3]     Jian-Miin Liu, cond-mat/0112084
[4]     Jian-Miin Liu, Chaos Solitons&Fractals, 12, 1111 (2001)
[5]     D. D. Clayton, Principles of Stellar Evolution and Nucleosythesis, University of Chicago Press, Chicago (1983)
[6]     R. D'E. Atkinson and F. G. Houtermans, Zeits. Physik, 54, 656 (1929)
[7]     G. Gamow, Phys. Rev., 53, 595 (1938)
[8]     Jian-Miin Liu, Modification of special relativity and its implications for the divergence problem in quantum field theory and the solar neutrino problem in standard solar models, to be published
[9]     J. N. Bahcall, Neutrino Astrophysics, Cambridge University Press, Cambridge (1989)
[10]    S. Turck-Chieze et al, Phys. Rep., 230, 57 (1993)
[11]    J. N. Bahcall et al, astro-ph/0010055
[12]    G. Shaviv, astro-ph/0010152
[13]    G. Fiorentini, B. Ricci and F. L. Villante, astro-ph/0011130